\documentclass[12pt]{article}

\def\half{{1\over 2}}
\def\bea{\begin{eqnarray}}
\def\eea{\end{eqnarray}}
\def \t{\tilde t}

\def\g {{1 \over \sqrt{1-{u^2 \over c^2 }}}}

\def\Dslash{\slash \negthinspace \negthinspace \negthinspace \negthinspace  D}

\usepackage{amsfonts}
\usepackage{amssymb}
\usepackage{epsfig}

\newcommand{\mop}[2]{\mathop{#1}^{#2}{}}
\newcommand{\mbb}{\mathbb}
\newcommand{\beqn}{\begin{eqnarray}}
\newcommand{\eeqn}{\end{eqnarray}}
\newcommand{\non}{\nonumber \\}

\newcommand{\cd}{{\cal D}}

\newcommand{\cn}{{\cal N}}

\newcommand{\cl}{{\cal L}}
\newcommand{\cf}{{\cal F}}

\newcommand{\co}{{\cal O}}
\newcommand{\cs}{{\cal S}}

\newcommand{\ct}{{\cal T}}

\newcommand{\tr}{{\rm tr}}
\newcommand{\FI}{{Fayet-Iliopoulos }}
\newcommand{\bb}[2]{\!\left[^{#1}_{#2}\right]\!}

\newcommand{\be}{\begin{equation}}
\newcommand{\ee}{\end{equation}}

\def\be{\begin{equation}}
\def\ee{\end{equation}}
\def\bea{\begin{eqnarray}}
\def\eea{\end{eqnarray}}
\def\ba{\begin{array}}
\def\ea{\end{array}}
\def\bd{\begin{displaymath}}
\def\ed{\end{displaymath}}

\def\tr{{\rm tr}}

\def\unit{1 \hskip-.3em \raise2pt\hbox{$ \scriptstyle |$ } }
\def\a{\alpha}
\def\b{\beta}

\def\d{\delta}
\def\e{\epsilon}           % Also, \varepsilon
\def\g{\gamma}

\def\k{\kappa}                    % Also, \varkappa (see below)
\def\l{\lambda}
\def\m{\mu}
\def\n{\nu}
\def\o{\omega}  
\def\th{\theta}                  %     \vartheta
\def\r{\rho}                                     %     \varrho
\def\s{\sigma}                                   %     \varsigma
\def\t{\tau}

\def\G{\Gamma}

\def\O{\Omega}

\def\half{{1 \over 2}}

\def\bop#1{\setbox0=\hbox{$#1M$}\mkern1.5mu
        \vbox{\hrule height0pt depth.04\ht0
        \hbox{\vrule width.04\ht0 height.9\ht0 \kern.9\ht0
        \vrule width.04\ht0}\hrule height.04\ht0}\mkern1.5mu}
\def\Box{{\mathpalette\bop{}}}                        % box
\def\pa{\partial}                              % curly d
\def\>{\rangle} %right angle
\def\<{\langle} %left angle
\def\Dsl{D \hskip-.6em \raise1pt\hbox{$ / $ } }
\def\to{\rightarrow}

\def\pa{\partial}

\def\+{\oplus}

\def\half{{1 \over 2}}

\def\lab{\label}
\def\sq2{\sqrt{2}}

\def\ba{\bar{\a}}
\def\bb{{\bar{\b}}}
\def\bg{\bar{\g}}
\def\bd{\bar{\d}}
\def\pb{\bar{\psi}}

\def\bj{\bar{\jmath}}

\begin{document}
%%%%%%%%%%%%%%%%%%%%%%%%%%%%%%%%%%%%%%%%%%%%%

\thispagestyle{empty}
\begin{flushright}
\vspace{-3cm}
{\small MIT-CTP-3683 \\
        hep-th/0509217}
\end{flushright}
\vspace{.5cm}

\begin{center}
{\bf\LARGE K\"ahler Anomalies in Supergravity \\[.3cm] and Flux Vacua
}

\vspace{1.5cm}

{\bf Daniel Z.~Freedman}$^{1,2}$ {\bf\hspace{.1cm}
and\hspace{.2cm} Boris K\"ors}$^{2,3,4}$ \vspace{1cm}

{\it
$^1$Department of Mathematics, Massachusetts Institute of Technology \\
Cambridge, Massachusetts 02139, USA \\

$^2$Center for Theoretical Physics, Laboratory for Nuclear Science \\
and Department of Physics, Massachusetts Institute of Technology \\
Cambridge, Massachusetts 02139, USA \\

$^3$II. Institut f\"ur Theoretische Physik der Universit\"at Hamburg \\
Luruper Chaussee 149, D-22761 Hamburg, Germany\\

$^4$Zentrum f\"ur Mathematische Physik, Universit\"at Hamburg \\
Bundesstrasse 55, D-20146 Hamburg, Germany

}

\vspace{1cm}

{\bf Abstract}
\end{center}\vspace{-.3cm}
We review the subject of K\"ahler anomalies in gauged
supergravity, emphasizing that field equations are inconsistent
when the K\"ahler potential is non-invariant under gauge
transformations or when there are elementary Fayet-Iliopoulos
couplings. Flux vacua solutions of
string theory with gauged $U(1)$ shift symmetries appear to avoid
this problem. The covariant K\"ahler anomalies
involve tensors which are composite functions of the scalars as
well as the gauge field strength and space-time curvature tensors.
Anomaly cancellation conditions will be discussed in a sequel to
this paper.

\vspace{-.5cm}

\clearpage

%%%%%%%%%%%%%%%%%%%%%%%%%%%%%%%%%%%%%%%%%%%%%%%%%%%

\section{Introduction}

This paper is devoted to the subject of K\"ahler anomalies in
gauged $\cn =1, ~D=4$ supergravity theories. The subject is
certainly not new, but we revisit it because gauged supergravity
appears as the effective four-dimensional theory in flux
compactifications of string theory and in recent applications of
Fayet-Iliopoulos couplings.\footnote{For early supergravity models in which the 
relevant structure appears see \cite{dzf1,drei,cast}.}
K\"ahler anomalies are not always
physically significant, but when they are the field equations of
the theory become inconsistent. The relation between anomalies and
inconsistency is emphasized in our work.

The models we consider contain the supergravity multiplet
$(e^i_\m,\Psi_\m)$, coupled to gauge multiplets $(A^a_\m, \l^a)$
and chiral multiplets $(z^\a, \psi^\a).$ The dynamics of the
chiral multiplets is that of a non-linear $\s$-model whose target
space is an $n$-dimensional K\"ahler manifold called $\ct$. In
these theories the (Majorana) gravitino covariant derivative
contains both the spin connection and a $U(1)$ axial gauge
connection, i.e.
\beqn
D_\mu \Psi_\n = \Big(\pa_\mu + \frac{1}{4}\o_{\m ij}
\g^{ij} + \half i B_\m \g_5\Big) \Psi_\n\ .
\eeqn
The K\"ahler connection $B_\m$ is
typically a composite function of the scalars $z^\a,\,  z^{\ba}$ and
the elementary vectors $A^a_\m$. The structure of
supergravity requires the K\"ahler connection to couple to the
gravitino and all other fermion fields.

The K\"ahler connection and the conventional axial couplings of
the fundamental gauge field $A^a_\m$ lead to anomalies of gauge
currents which we explore through their effects on the consistency
condition for gauge field equations of motion\footnote{Gauge group
indices are denoted by superscripts in the text and by
``overscripts" in equations.}
\be \lab{gfeq}
D_\m \mop Fa\!^{\m\n} =\mop Ja\!^{\n}.
\ee
The left side vanishes identically if one applies a further $D_\n$,
\be \lab{conc}
0 \equiv D_\n D_\m \mop Fa\!^{\m\n} = D_\n \mop Ja\!^{\n},
\ee
and the current must be conserved for consistency. If the
classical action is gauge invariant, then the current is conserved
classically for field configurations which satisfy classical
equations of motion. But, as is very well known, fermion triangle
anomalies, which yield (schematically)
\be
D_\n \mop Ja\!^{\n} ~\propto~ \e^{\m\n\r\s}\, \tr
\big[ \mop Ta F_{\m\n}F_{\r\s}\big] \, ,
\ee
spoil current conservation and consistency at the quantum level
\cite{gj}.

It is the consistency condition (\ref{conc}) that we study in
supergravity models. By classical manipulation we express $ D_\n
J^{a\n}$ in terms of bilinear fermion currents, and we evaluate
their anomalies, using the Fujikawa method \cite{fuji} to express
results in terms of covariant anomalies. The analysis is a quite
complicated affair in the supergravity models, so we develop the
basic ideas in simpler truncated models in section 3 before
applying them to the general situation in section 4.  This follows
a review of gauged K\"ahler non-linear $\s$-models in section 2.

In section 5, we turn our attention to supergravity models which
descend from flux compactifications of superstring theory. We
study a generic model with gauge shift symmetry and show that the
field equations are consistent although there are uncanceled
triangle anomalies. The question of anomaly cancellation in other
models then remains. For gauge current anomalies this requires the
conversion of covariant to consistent anomalies. This is done in a
class of models which include Fayet-Iliopoulos couplings in a
sequel to this paper \cite{EFK}.

%%%%%%%%%%%%%%%%%%%%%%%%%%%%%%%%%%%%%%%%%%%%%%%%%%%

\section{K\"ahler manifolds and holomorphic isometries}
\setcounter{equation}{0}

We begin by reviewing (and defining notation for) the local geometry of
K\"ahler manifolds and their continuous isometries. For more information, see
\cite{bag,bonn,fag}. We also discuss K\"ahler anomalies in general terms.

The scalar fields are complex coordinates on $\ct$. They are
denoted collectively by $z^A,~ A =1,\ldots, 2n$, which split into
$n$ holomorphic coordinates $z^\a$ and $n$ anti-holomorphic
$z^{\bar{\a}}$. The metric splits in the standard fashion
\be \lab{kmet}
ds^2 = G_{AB}dz^A dz^B = 2G_{\a\bar{\b}}dz^\a
dz^{\bar{\b}}\ ,
\ee
in which the metric tensor can be expressed as second derivative of the
K\"ahler potential $K(z,\bar{z})$, viz.
\be \lab{kpot}
G_{\a\bar{\b}} = K,_{\a\bar{\b}} = \frac{\pa^2}{\pa z^\a \pa z^\bb} K(z,\bar{z}) \ .
\ee
We use a comma to denote partial derivatives,
and a semi-colon for K\"ahler covariant
derivatives, e.g $V^\a;_{\b} = V^\a,_\b + \G^\a_{\b\g}V^\g$. The
only non-vanishing components of the Christoffel connection
$\G^A_{BC}$ are the all-holomorphic $\G^\a_{\b\g}= G^{\a
\bd}G_{\g\bd},_\b$ and its complex conjugate $\G^{\ba}_{\bb\bg}$.
The curvature tensor $R_{AB}{}^C{}_D$ enjoys the usual symmetries,
but the only non-vanishing components are $R_{\a\bb}{}^\g{}_\d = -
\G^\g_{\a\d},_{\bb}$ and those related by symmetry and complex
conjugation.

It is significant that
the metric $G_{\a\bb}$ does not change under K\"ahler transformations
of the potential
\be \lab{ktrf}
K(z,\bar{z}) ~\to~ K^\prime(z,\bar{z})=K(z,\bar{z})+f(z)
+\bar{f}(\bar{z})\ .
\ee
In general the potential is not a global scalar, but is locally
defined in each coordinate chart, the definitions in overlapping
charts related by a K\"ahler transformation. Most terms in gauged
supergravity Lagrangians are invariant under (\ref{ktrf}), but the
K\"ahler connection changes by $B_\m \to B_\m + \pa_\m {\rm
Im}(f(z))$. Classically this can be compensated by an axial gauge
transformation of the fermions, e.g. $\Psi_\m \to {\rm
exp}(-\frac{i}{2}{\rm Im}(f(z)) \g_5) \Psi_\m$, but this
transformation is anomalous. This is the basic K\"ahler anomaly.
It does not necessarily make the theory inconsistent at the
quantum level. In this paper we focus on situations in which a
gauge transformation of the bosonic fields induces the K\"ahler
transformation (\ref{ktrf}). The accompanying fermion gauge
transformation is anomalous and the theory does become
inconsistent. We now review the machinery needed to implement
gauge symmetry.

In our physical models gauge fields couple to holomorphic
isometries of $\ct$. Each such isometry is defined by a
holomorphic Killing vector $X^{a\a}(z)$. There is a real scalar
Killing potential $D^a(z,\bar{z})$ related to each $X^a_\a$ by
$D,_{\a}^{\!\! a} = -i X^a_\a$.  The holomorphic Killing vectors
satisfy $X^a_{\a;\bb} + X^a_{\bb;\a}=0$ and 
generate a Lie algebra via the Lie bracket relations
\be \lab{lieb}
\mop Xa\!^\b \mop Xb\!^\a,_{\b} - \mop Xb\!^\b \mop Xa\!^\a,_{\b} = f^{abc} \mop Xc\!^\a\ .
\ee
We assume that this Lie algebra is a direct sum of a compact
simple algebra and possible $U(1)$ subalgebras. The Killing
potentials in the non-abelian simple sector are uniquely defined
by the requirement that they transform in the adjoint
representation, i.e.
\be \lab{dadj}
\mop Xa\!^\a\! \mop Db\!,_{\a} + \mop Xa\!^{\ba}\! \mop Db\!,_{\ba} = f^{abc}
\mop Dc\ ,
\ee
while those in abelian factors are defined up to an additive
integration constant, i.e., $D^a \to D^a + \xi^a$, where $\xi^a$
is a Fayet-Iliopoulos coupling in the physical context.

The Killing vector relations above state the fact that the
K\"ahler metric $G_{\a\bb}$ is invariant under the isometry, and
the connection and curvature tensor are also invariant. However,
it is important to note that the K\"ahler potential need not be
invariant. It must satisfy only the weaker condition
\be \lab{ktrf2}
\mop Xa\!^\a K,_{\a} + \mop Xa\!^{\ba} K,_{\ba} =\mop Fa(z) + \mop {\bar{F}}a(\bar{z})
\ee
which implies that the metric is invariant. The holomorphic
quantity $F^a(z)$ will play an important role in our discussion.
It is easy to show that $D^a$ and $F^a$ are related by
\be \lab{df}
\mop Da = i\big( K,_{\a} \mop Xa\!^\a - \mop Fa \big) =  -i\big(
K,_{\ba} \mop Xa\!^{\ba} - \mop {\bar{F}}a \big)\ .
\ee
Note that even in the case of an invariant K\"ahler potential
(\ref{ktrf2}) always leaves the freedom to add an imaginary
constant to $F^a$. In the non-abelian case, the value of that
constant is fixed by imposing (\ref{dadj}) while for an abelian
isometry it reflects the freedom to add a Fayet-Iliopoulos
coupling $\xi^a$ to the theory. Even when the K\"ahler potential
is gauge invariant and the right side of (\ref{ktrf2}) vanishes,
there are K\"ahler anomalies which threaten the consistency of the
supergravity theory if $\xi^a={\rm Im}(F^a)\neq 0$.

In gauged supergravity all fermion fields couple to the composite
K\"ahler connection which is given by
\be \lab{kcon}
B_\m = \frac{1}{2i}\Big( K,_{\a}D_\m z^\a + \mop Aa\!_\m \mop Fa -
    K,_{\ba}D_\m z^{\ba} - \mop Aa\!_\m \mop{\bar{F}}a \Big)
=\frac{1}{2i}\Big( K,_{\a}\pa_\m z^\a  -
    K,_{\ba}\pa_\m z^{\ba}  \Big) + \mop Aa\!_\m \mop Da
    \ .
\ee
It couples with gravitational strength and thus appears with
coefficient $\k^2$ which is set to $\k=1$ in the present notation.
Under the gauge transformation
\beqn\lab{varbos}
\d z^\a = \mop \theta a \mop Xa\!^{\a}\ ,\quad
\d {z}^{\bb} = \mop \theta a \mop Xa\!^{\bb}\ ,\quad
\d \mop Aa\!_\m ~=~ \pa_\m \mop \theta a + f^{abc} \mop Ab\!_\m \mop \theta c \ ,
\eeqn
one may show that $B_\m$ transforms as an abelian gauge connection, that is
\be \lab{bgau}
\d B_\m = \pa_\m (\mop \theta a {\rm Im}(\mop Fa) )\ .
\ee
%
%Another property of $B_\m$ useful in the computations below is
%
%\bea \lab{bprop}
%\frac{\d B_\m}{\d A^a_\n} &=& \mop Da \d^\n_\m
%\ ,
%\non
%\frac{\d B_\m}{\d z^\a} &=& + i G_{\a \bb}D_\m z^{\bb} +\frac{i}{2}K,_\a
%\pa_\m \ ,
%\non
%\frac{\d B_\m}{\d z^{\bb}} &=& -i G_{\a \bb}D_\m z^{\a} - \frac{i}{2}K,_\a \pa_\m
%\ .
%\eea
%
%The last two expressions are Euler variations in which the last
%$\pa_\m$ acts on other factors in the expression which is varied.
%

The mathematical background just described leads to the bosonic
Lagrangian of the non-linear $\s$-model on $\ct$
\be \lab{lbos}
\cl_b = -G_{\a\bb} D_\m z^\a D^\m z^{\bb}
\ee
with covariant derivatives
\bea \lab{dbos}
D_\m z^\a &=& \pa_\m z^\a - \mop Aa\!_\m \mop Xa\!^{\a}\ , \quad
D_\m z^{\bb} ~=~ \pa_\m z^{\bb} - \mop Aa\!_\m \mop Xa\!^{\bb}\ .
\eea

The supersymmetric partners of the bosons are Weyl spinors which
transform as tangent vectors under holomorphic diffeomorphisms of
$\ct$. We write these as 4-component spinors with chiral projectors $L,R
= (1 \mp \g_5)/2$. Thus a chiral multiplet consists of the set
$(z^\a, L\psi^\a)$, while an anti-chiral multiplet is $(z^{\bb},
\bar{\psi}^{\bb}R)$. The fermion kinetic Lagrangian is
\bea \lab{lfer}
\cl_f ~=~ G_{\a\bb} \bar{\psi}^{\bb} \g^\m D_\m L \psi^\a
~=~ -  G_{\a\bb} (D_\m\bar{\psi}^{\bb}R) \g^\m \psi^\a
\eea
with covariant derivatives\footnote{\lab{foot1}The K\"ahler connection $B_\mu$ will
be included in later sections. The Lagrangian of the supersymmetric $\s$-model
\cite{zum} includes another term
$\frac{1}{4}R_{\a\bb\g\bd}(\pb^\a L\psi^\g)(\pb^{\bb}R\psi^{\bd})$
and a Yukawa coupling to gauginos.
We will discuss this quartic term in footnote \ref{foot2}, and the
Yukawa coupling will be included in the supergravity analysis.}
\bea \lab{dpsi}
D_\m L\psi^\a &=& (\pa_\m \d^\a_\b + \G^\a_{\b\g}\pa_\m z^\g -
\mop Aa\!_\m \mop Xa\!^{\a};_{\b}) L\psi^\b \ ,
\non
D_\m\bar{\psi}^{\bb}R &=& (\pa_\m \d^{\bb}_{\bg} +
\G^{\bb}_{\bg\bd}\pa_\m \bar{z}^{\bd} - \mop Aa\!_\m \mop
Xa\!^{\bb};_{\bg})\pb^{\bg}R\ .
\eea
It is a useful exercise to show that under the gauge
transformations of (\ref{varbos}) for the bosons and
\be  \lab{gau}
%\d z^\a &=& \mop \theta a \mop Xa\!^{\a}\ ,
%\quad
\d L\psi^\a ~=~ \mop \theta a \mop Xa\!^{\a},_{\b} L\psi^\b \ ,
\quad
%\d \mop Aa\!_\m ~=~ \pa_\m \mop \theta a + f^{abc} \mop \theta b \mop Ac\!_\m ,
\ee
for fermions the covariant derivatives transform as (holomorphic components of)
tangent vectors, i.e.
\bea \lab{gaud}
\d D_\mu z^\a &=& \mop \theta a \mop Xa\!^\a,_{\b} D_\m z^\b\ ,
\quad
\d D_\m L\psi^\a ~=~ \mop \theta a \mop Xa\!^\a,_{\b} D_\m L\psi^\b\ .
\eea
%

%%%%%%%%%%%%%%%%%%%%%%%%%%%%%%%%%%%%%%%%%%%%%%%%%%%%%%%%%%%%%%%%%%%%%%%%%%%%%%%

\section{Anomalies and Inconsistency}
\setcounter{equation}{0}
\lab{an+inc}

In this section we will derive the consistency condition in a
succession of models which gradually incorporate the features of
the full gauged supergravity models which we tackle in section 4.
It is worth stating our strategy in general terms before embarking
on the detailed analysis.

\subsection{The strategy}

The models considered include gauge fields $A_\m^a$, scalars
$z^\a,~ z^{\ba}$ of a K\"ahler $\s$-model, and fermions
(the gravitino, gauginos and chiral fermions) which we temporarily denote by $\psi$. The gauge current
in all models is defined by
\beqn \lab{curdef}
\mop Ja\!^{\r} \equiv - \frac{\d (\cl+\frac14 F^b_{\m\n}F^{b\m\n})}{\d A^{a\r}}
= \mop Ja\!^{\r}_b + \mop Ja\!^{\r}_f
\eeqn
and is the sum of a purely bosonic term $J^{a\n}_b$ and a term
$J^{a\n}_f$ involving the fermions. The gauge field satisfies the
Yang-Mills equation (\ref{gfeq}) and the consistency condition
(\ref{conc}) requires conservation of the full current, i.e.
\be \lab{cona}
D_\n(\mop Ja\!^{\n}_b +\mop Ja\!^{\n}_f) = 0\ .
\ee
One way to isolate the anomalous fermion terms in the divergence
of the current is to use the scalar equations of motion (which
include fermion terms) to evaluate $D_\n J^{a\n}$ in (\ref{cona}).
However, this process is rather complicated in the context of the
gauged $\s$-model, and it can be bypassed as we now describe.

In any gauge invariant model there is a functional identity which
expresses the gauge invariance at the {\it classical} level, namely
\bea \lab{gi1}
\mop \theta a (x)\Big[
D_\n \mop Ja\!^{\n} +\mop Xa\!^\a \frac{\d \cl}{\d
z^\a}
+\mop Xa\!^{\bb}\frac{\d \cl}{\d z^{\bb}} + \mop\d a \pb \frac{\d \cl}{\d \pb}\Big] ~\equiv~ 0\ ,
\eea
in which the gauge variations of the bosons of (\ref{varbos})
appear and gauge variations of all the fermions are symbolically
denoted by $\d^a\psi$. As is very well known, this gauge identity
tells us that, no matter how complicated the model, $D_\n \mop
Ja\!^{\n}$ vanishes {\it classically} if {\it all} charged fields
satisfy their equations of motion. It may also be interpreted as
the statement that, if the {\it scalar} equations of motion are
satisfied, the functional form of $D_\n \mop Ja\!^{\n}$ is the
negative of the fermion gauge variation of the Lagrangian. The
consistency condition
(\ref{conc}) may be viewed as a cancellation condition for the
correction to the gauge identity due to anomalies.

%%%%%%%%%%%%%%%%%%%%%%%%%%%%%%%%%%%%%%%%%%%%%%%%%%%%%%%%%%%%%%%%%%%%%%%%%

\subsection{A model with gauginos}

We now apply the strategy just outlined to derive the consistency
condition for a model which incorporates some features of gauged
supergravity theories. The model contains a (non-abelian) gauge field coupled
to complex scalars $z^\a$ which determine a gauged non-linear
$\s$-model whose target space is the $n$-dimensional K\"ahler
manifold $\ct$ and to Majorana fermions $\l^a$ in the adjoint
representation of the gauged isometry group $G$. Since we leave out
the chiral fermions, the superpartners of the $z^\a$, this model is
not supersymmetric. Its Lagrangian density is
\be \lab{toy2}
\cl = -\frac14 \mop Fa\!_{\m\n}\mop Fa\!^{\m\n} - G_{\a\bb} D_\m z^\a D^\m z^{\bb}
+ \half \mop {\bar{\l}}a \g^\m D_\m \mop \l a\ .
\ee
The boson covariant derivatives are given in (\ref{dbos}), while
the fermion covariant derivative
\be \lab{dlam}
D_\m \mop \l a = \pa_\m \mop \l a + f^{abc} \mop Ab\!_\m \mop \l c +\half i B_\mu \g_5 \mop \l a
\ee
includes both the expected elementary connection and the composite
K\"ahler connection $B_\m$ defined in (\ref{kcon}).
We include $B_\m$ in the present non-gravitational model to illustrate the important
role it plays in the anomaly analysis.

The Lagrangian (\ref{toy2}) is gauge invariant under the
transformations of (\ref{gau}) for bosonic fields and
\be \lab{gaulam}
\d \mop \l a = f^{abc}\mop \l b \mop \theta c - \half i\mop \theta b
{\rm Im}(\mop F b) \g_5 \mop \l a
\ee
for the fermions. It also has the global axial symmetry
$\d \l^a = i \a\g_5\l^a$ with Noether current
\be \lab{noe}
N^\m = - \frac{i}{2} \mop{\bar{\l}}a \g^\m\g_5\mop\l a\ .
\ee
Although we do not need the explicit form of the equations of
motion, we record them for completeness. One can write the gauge
field equation as
%\footnote{We use the $2n$-valued indices $A\, ,B$
%to simplify the expression. They independently take holomorphic
%and anti-holomorphic values, i.e.\ $A\in\{\a,\, \ba\}$, etc.}
%
\bea \lab{aeq}
D_\m \mop Fa\!^{\m\n} = \mop Ja\!^{\n} \equiv
-\frac{\d(\cl+\frac14 F^b_{\r\s} F^{b\r\s})}{\d A^a_\n} %\non
= -
G_{AB} \mop Xa\!^{A}D^\n z^B +
\mop ja\!^{\n} + \half \mop Da N^\n
\eea
with the adjoint gauge current $j^{a\n} = \half 
f^{abc}\bar{\l}^b\g^\n\l^c$. The equation for the scalar $z^\a$ is
\bea \lab{zeq}
D^\n D_\n z^\a
%&\equiv& ( \pa^\n \d^\a_\b + \G_{\b\g}^\a \pa^\n
%z^\g
%- \mop Aa\!^\n \mop Xa\!^\a;_\b)  D_\n z^\b
%\non
&=&
- \half i \Big(N^\n D_\n z^\a +
\frac12 G^{\a\bb}K,_{\bb}\pa_\n N^\n \Big)
%\ ,
%\non
%D^\n D_\n z^{\bb}&=& + \half i \Big(N^\n D_\n z^{\bb} +
%\frac12 G^{\a\bb}K,_{\a}\pa_\n N^\n\Big)
\ .
\eea
The left sides of these equations contain the gauge covariant
harmonic map operator $D^\n D_\n z^\a \equiv ( \pa^\n \d^\a_\b +
\G_{\b\g}^\a \pa^\n z^\g
- \mop Aa\!^\n \mop Xa\!^\a;_\b)  D_\n z^\b$. The equation for
$z^{\bar\a}$ is the complex conjugate of (\ref{zeq}).

To derive the consistency condition we insert the Dirac conjugate
of the fermion gauge variation (\ref{gaulam}) in the gauge
identity (\ref{gi1}), assuming, of course, that
(\ref{zeq}) is satisfied. After some index shuffling, we obtain
\be \lab{gi2}
\mop \theta a D_\n \mop Ja\!^{\n} =
\mop \theta a \Big [ f^{abc} \mop{\bar{\l}}b \g^\m D_\m\mop\l c + \half i\, {\rm Im}(\mop Fa) \mop{\bar{\l}}b \g_5\g^\m  D_\m \mop\l b\Big ].
\ee
Note that the $B_\m$ connection cancels in $ D_\m\l^b$ due to the
symmetry properties of Majorana bilinears. The Majorana properties
also allow us to extract $D_\m$ as a total gauge covariant
derivative. Then dropping the overall factor of $\theta^a$, we can
express the consistency condition as
\be \lab{anom1}
0 =D_\n \mop Ja\!^{\n}= D_\n \mop ja\!^{\n}\, + \half {\rm Im}(\mop Fa) \pa_\n N^\n \ .
\ee
The currents $N^\n$ and $j^{a\n}$ are conserved classically, so
we would obtain the identity $0=0$
reflecting the classical gauge invariance of the theory. The
advantage of the consistency condition (\ref{anom1}) is that it
enables us to bring in the quantum level violation of gauge
invariance due to the anomaly.

%%%%%%%%%%%%%%%%%%%%%%%%%%%%%%%%%%%%%%%%%%%%%%%%%%%%%%%%%%%%%%%%%%%%%%%%%5

\subsection{The axial anomaly with gauginos}
\lab{secgaug}

The anomalous contribution of (\ref{anom1}) may be
calculated by the method of Fujikawa in which we consider the
change in the fermion measure in the path integral due to the
changes of variable $\d\l^a = i \a (x) \g_5 \l^a$ and $\d\l^a =
f^{abc} \l^b \th^c(x)$. In order to minimize repetition of well known
material we will give only brief discussion and refer readers to
the monograph of Fujikawa and Suzuki \cite{fuji}. Results for the
several cases of anomalies needed in this paper can be obtained by
modification of the appropriate sections of \cite{fuji}. All
results are stated in terms of the {\it covariant} anomaly.

In particular the anomalous divergence $\pa_\n N^\n$ in our model
may be obtained by rewriting the Majorana kinetic action in terms
of Weyl spinors and using sections 6.4 and 6.4.1 of \cite{fuji}.
Fermion mode functions are defined to satisfy
\be \lab{mode1}
\Dslash_R\, \Dslash_L \mop \phi a\!_n = - \r_n^2 \mop \phi a\!_n
\ee
with
\bea \lab{mode2}
\mop \Dslash{ab}\!_L &=& \g^\m \mop D{ab}\!_\m L
 ~=~ \g^\m\Big[ (\pa_\m - \half i B_\m)\d^{ab} - f^{abc}\mop Ac\!_\m\Big
]L\ ,
\non
\mop \Dslash{ab}\!_R &=& -\mop \Dslash{ab}\!_L^\dagger ~=~
 \g^\m\Big[ (\pa_\m - \half i B_\m)\d^{ab} - f^{abc}\mop Ac\!_\m\Big]
 R\ .
\eea
Note that
\be \lab{d2}
\Dslash_R\, \Dslash_L = \Big[ D^\m D_\m +\half\g^{\m\n}\cf_{\m\n}\Big] L\ ,
\ee
in which $\cf^{ab}_{\m\n}$ is the field strength of the full
connection in (\ref{mode2}), namely
\bea \lab{f1}
\mop \cf{ab}\!_{\m\n} &=& - f^{abc} \mop Fc\!_{\m\n} -\half i B_{\m\n}\d^{ab} \ ,
\non
\mop Fc\!_{\m\n} &=& \pa_\m \mop Ac\!_\n -  \pa_\n \mop Ac\!_\m + f^{cde} \mop Ad\!_\m \mop Ae\!_\n \
,\quad  B_{\m\n} ~=~ \pa_\m B_\n - \pa_\n B_\m\ .
\eea
The anomaly of the Noether current $N_\m$ of $\d\l^a = i\a(x)
\g_5\l^a$ is
\be  \lab{axanom}
\langle \pa_\n N^\n \rangle ~=~
- \frac{1}{32\pi^2} \e^{\m\n\r\s}\mop \cf{ab}\!_{\m\n}\mop
\cf{ba}\!_{\r\s} ~=~\frac{1}{32\pi^2} \e^{\m\n\r\s}
\Big[ C_2(G)
\mop Fb\!_{\m\n}\mop Fb\!_{\r\s} +
\frac{1}{4} n_\l B_{\m\n}B_{\r\s}\Big]\ ,
\ee
where $n_\l= {\rm dim}(G)$ is the total number of gauginos and
$C_2(G)\d^{ab} = f^{acd}f^{bcd}$ is the adjoint Casimir operator.
The vector current is also anomalous here because it couples to the $B_\m$ connection.
Its anomaly is
\beqn \lab{gauganom}
\< D_\n \mop ja^{\n} \>~=~ \frac i{32\pi^2} \e^{\m\n\r\s} f^{abc}
\mop \cf{cd}\!_{\m\n}\mop \cf{db}\!_{\r\s}
~=~ \frac 1{32\pi^2} C_2(G) \e^{\m\n\r\s} \mop F{a}\!_{\m\n} B_{\r\s} \
.
\eeqn
Combining (\ref{axanom},\ref{gauganom}), and (\ref{anom1}), we find that
the ``inconsistency condition'' of the model reads
\be \lab{incon1}
0 ~=~ \frac12 {\rm Im}(\mop Fa)\, \e^{\m\n\r\s}\Big[ C_2(G)
\mop Fb\!_{\m\n}\mop Fb\!_{\r\s} +
\frac{1}{4} n_\l B_{\m\n}B_{\r\s}\Big] +
C_2(G) \e^{\m\n\r\s} \mop F{a}\!_{\m\n} B_{\r\s}
\ .
\ee
The model appears to be fatally inconsistent since the
coefficients of all three terms on the right-hand-side have to
vanish. We will discuss in \cite{EFK} the possibility that parts
of the anomaly may be removed by adding local non-gauge invariant
polynomials in the gauge potential and the K\"ahler connection to
the classical action. However, it will turn out that no such
counter terms exist to remove the term involving
$B_{\m\n}B_{\r\s}$. Therefore, the model is indeed inconsistent unless
the K\"ahler potential is gauge invariant. 

%%%%%%%%%%%%%%%%%%%%%%%%%%%%%%%%%%%%%%%%%%%%%%%%%%%%%%%%%%%%%%%%%%%%%%%%%%%%

\subsection{Models with chiral fermions}

We now analyze another model in similar fashion, a model in which
the non-abelian gauge field is coupled both to the bosons and
fermions of the K\"ahler manifold $\s$-model. The features of both
models discussed in this section, and more, will be combined to
treat the general gauged supergravity model in section \ref{full}.
The Lagrangian of this model is
\be\lab{cfml}
\cl =  -\frac14 \mop Fa\!_{\m\n}\mop Fa\!^{\m\n} -G_{\a\bb} D_\m z^\a D^\m z^{\bb}
+  G_{\a\bb} \bar{\psi}^{\bb} \g^\m D_\m L \psi^\a
\ee
with fermion covariant derivative\footnote{Note that $B_\m$ couples to
$L\psi^\alpha$ and $L\l^a$ with opposite sign.}
\be \lab{dpsi2}
D_\m L\psi^\a = \Big( \pa_\m \d^\a_\b + \G^\a_{\b\g}\pa_\m z^\g -
\mop Aa\!_\m \mop Xa\!^{\a};_{\b}+ \half i B_\m \d^\a_\b\Big) L\psi^\b\ .
\ee
The Lagrangian is gauge invariant if the boson transformations of
(\ref{gau}) are combined with
\be \lab{gau1}
\d L\psi^\a = \mop \theta a \Big( \mop Xa\!^{\a},_{\b} L\psi^\b
 - \frac{i}{2}{\rm Im}(\mop Fa) L\psi^\a\Big) \ ,
\ee
and there is a global axial symmetry $\d L\psi^\a = i\a L\psi^\a$ with
Noether current
\be \lab{noe2}
N^\m = -i G_{\a\bb}\bar{\psi}^{\bb}\g^\m L \psi^\a\ .
\ee
The gauge field equation is
\bea \lab{aeq2}
D_\m \mop Fa\!^{\m\n} &=& \mop Ja\!^{\n} ~\equiv~
-\frac{\d(\cl+\frac14 F_{\r\s}^bF^{b\r\s})}{\d A^a_\n}~=~ \mop
Ja\!^{\n}_b + \mop Ja\!^{\n}_f + \half \mop Da N^\n\ ,
\non
\mop Ja\!^{\n}_b &=& - G_{\a\bb} \Big( \mop Xa\!^{\a} D^\n z^\bb + \mop Xa\!^{\bb} D^\n z^\a \Big) \ ,\quad
\mop Ja\!^{\n}_f ~=~ \mop Xa\!_{\bb;\a} \pb^{\bb}\g^\n L \psi^\a\ ,
\eea
and the scalar equations are
\bea \lab{zeom}
\frac{\d \cl}{\d z^{\bg}} &=& 0 ~=~  G_{\a\bg} D_\m D^\m z^\a
  +R_{\d\bb\a\bg}D_\n z^{\d}(\pb^{\bb}\g^\n L\psi^\a)
\\
&& \hspace{2.5cm}
+ G_{\a\bb,\bg}\pb^{\bb}\g^\m D_\m L\psi^\a
 + \frac{i}{2}G_{\d\bg}D_\m z^\d N^\m + \frac{i}{4}K,_{\bg}\pa_\m N^\m
\ ,
\nonumber
\\  \lab{zbeom}
\frac{\d \cl}{\d z^{\g}} &=&  0 ~=~  G_{\g\bb} D_\m D^\m z^{\bb}
 -R_{\bd\a\bb\g}D_\n z^{\bd}(\pb^{\bb}\g^\n L\psi^\a) \\
&& \hspace{-1cm}
+ G_{\a\bb,\g}\pb^{\bb}\g^\m D_\m L\psi^\a - \frac{i}{2}G_{\g\bb}D_\m
z^{\bb} N^\m - \frac{i}{4}K,_{\g}\pa_\m N^\m
- G_{\d\bb}\G^\d_{\a\g}D_\m(\pb^{\bb}\g^\m L\psi^\a)\ .
\nonumber
\eea
Note that $R_{\bg\b\ba\d}= R_{\bg\d \ba\b}.$
The second equation becomes the conjugate of the first after applying
the covariant divergence to the fermions in the last term of
(\ref{zbeom}).

It is a very intricate task (which we have done) to derive the consistency
condition for this model by combining equations of motion. The
complicated scalar field equations are indicative of the difficulty. It is
far simpler to use the gauge identity (\ref{gi1}). For this we simply need
the fermion gauge variation of the Lagrangian, using both $\d L\psi^\a$ in
(\ref{gau1}) and its conjugate $\d \bar{\psi}^\bb R$. After dropping the common
factor $\theta^a$, the gauge identity can be written immediately
as the consistency condition\footnote{\label{foot2} Suppose that
we modify the model by adding the quartic $R \psi^4$ term in
footnote
\ref{foot1} that is part of the supersymmetric non-linear
$\s$-model and repeat the analysis. 
The consistency condition would be modified by the fermion gauge
variation of the quartic term. We will assume
that this does not modify the anomalies and leave this complication
aside.} 
\beqn \lab{gava}
D_\n \mop Ja\!^{\n} &=&
\\
&&\hspace{-1.5cm}
- G_{\a\bb}\Big[ \Big( \mop
Xa\!^{\bb},_{\bg}
+ \frac{i}{2} {\rm Im}(\mop Fa) \delta^\bb_{\bg}\Big)
 \pb^{\bg}\g^\m D_\m L\psi^\a -
\Big( \mop Xa\!^{\a},_\g - \frac{i}{2} {\rm Im}(\mop Fa) \delta^\a_{\g}\Big)
 (D_\m \pb^{\bb}R)\g^\m \psi^\g \Big]\ .
\nonumber
\eeqn
The non-covariant terms in this expression originate in the fact
that $\s$-model fermions transform under gauge variations as
tangent vectors on the target space and thus carry the
non-covariant factor $X^{a\a},_\b$. The result (\ref{gava}) may be
rearranged to read
\bea \lab{con4}
0 ~=~ D_\n \mop Ja\!^{\n} &=&
\frac12 \Big( G_{\g\bb} \mop Xa\!^{\g},_{\a} - G_{\a\bg} \mop Xa\!^{\bg},_{\bb} \Big)
D_\n(\pb^{\bb}\g^\n L\psi^\a) + \half {\rm Im}(\mop Fa) \pa_\n N^\n \\
&& - \frac12 \Big( G_{\g\bb} \mop Xa\!^{\g},_\a + G_{\a\bg} \mop Xa\!^{\bg},_\bb \Big)
(\pb^{\bb}\g^\m D_\m L\psi^\a-(D_\m \pb^{\bb})\g^\m L\psi^\a)\ .
\nonumber
\eea
We are again in a situation in which the consistency condition
would be satisfied if the classical fermion field equations are
used, but the form of the right side allows us to probe possible
quantum anomalies. Indeed, the anomalies of $ D_\n(\pb^{\bb}\g^\n
L\psi^\a)$ and $\pa_\n N^\n$ will be obtained in the next section.
One might suspect an anomalous one-loop contribution from the
fermion bilinear in the last term of (\ref{con4}) because this
term resembles the trace of a fermion stress tensor. However, we
have checked that the contributing Feynman diagrams conform to the
trace anomaly calculation of \cite{traceanom} and
cancel.\footnote{This is also described in section 19.5 and figure
19.10 of \cite{pesk}. The trace anomaly in Yang-Mills theory comes
from another diagram, not present in our case, involving the gauge
field stress tensor and fermion vacuum polarization.}

%%%%%%%%%%%%%%%%%%%%%%%%%%%%%%%%%%%%%%%%%%%%%%%%%%%%%%%%%%%%%%%%%%%%%%%%

\subsection{The axial anomaly with chiral fermions}

The first two of the three terms of (\ref{con4}) contain quantum
anomalies whose calculation can be modeled on that of section 6.4
of \cite{fuji}. To adapt the treatment of section 6.4 to the
situation of the non-linear $\s$-model we introduce frames on the
target space $\ct$ via
\be \lab{frame}
ds^2 = 2G_{\a\bb} dz^\a dz^{\bb} = 2 \d_{i\bj}e^i_\a dz^\a e^{\bj}_{\bb}dz^{\bb}
\ee
and use the frame basis to reexpress the $\s$-model fermions as
\be \lab{ffer}
L \psi^\a = e^\a_i L \psi^i\ ,\quad \pb^{\bb}R=
e^{\bb}_{\bj}\pb^{\bj}R \ ,
\ee
where $e^\a_i,~e^{\bb}_{\bj}$ are inverse frames. The fermion kinetic
Lagrangian of our model can be rewritten in the frame basis as
\be \lab{lfram}
G_{\a\bb} \bar\psi^{\bb} \gamma^\m D_\m L\psi^\a = \pb_i \g^\m D_\m L\psi^i
\ee
with
\bea  \lab{frstuff}
\pb_i R &=& \d_{i\bj}\pb^{\bj}R\ ,
\quad
\mop Xa\!^{i};_{j} ~=~ e^i_\a \mop Xa\!^{\a};_{\b}e^\b_j\ ,
\quad
\O^i_{Cj} ~=~ e^i_\b e^\b_{j,C} + e^i_\a\G^\a_{\b C}e^\b_j\ ,
\non
D^i_{\mu j}L\psi^j &=& \Big[ (\pa_\m +\half i B_\m)\d^i_j +\O^i_{C
j}\pa_\m z^C - \mop Aa\!_\m \mop Xa\!^{i};_{j}\Big] L\psi^j\ .
\eea
The $2n$-valued index $C$ appears because the K\"ahler spin connection
couples to both $\pa_\m z^\g$ and $\pa_\m z^{\bg}$.

As discussed in \cite{bny}, the fermions $\psi^i$ are sections of
a holomorphic vector bundle on $\ct$ with structure group $U(n)$.
The new fermion kinetic term (\ref{lfram}) is separately invariant under
diffeomorphisms of $\ct$ and unitary transformations of the frame
and fermion fields, viz.
\bea \lab{untrf}
e^i_\a &\to& U^i_j e^j_\a\ ,
\quad
\ e_{i\ba} = \d_{i\bar\imath}e^{\bar\imath}_{\ba} ~\to~ e_{j\ba}U^{\dagger j}_i\ ,
\non
\psi^i &\to& U^i_j \psi^j\ ,
\quad
\pb_i ~\to~ \pb_j U^{\dagger j}_i\ ,
\eea
where $U^i_j(z,\bar{z})$ is a unitary matrix. This means that we
can apply the discussion of section  6.4 of \cite{fuji} quite directly
and obtain the anomalous response to the transformation
(\ref{untrf}) of the fermion fields in the path integral measure.
For this purpose we introduce a standard basis of generators
$T^{ai}_j$, with $a = 0,1,\ldots ,n^2-1,$ of the fundamental
representation of $U(n)$, normalized by ${\rm
tr}(T^aT^b)=\frac12\d^{ab}$, and write $U=\exp(i\a^aT^a)$. The
$U(1)$ generator is $T^{0i}_j =
\d^i_j/\sqrt{2n}$.

Following \cite{fuji}, we expand the field $\psi^i$ in mode
functions $\phi^i_n$ which satisfy (\ref{mode1}), but with the new
operator
\be
(\Dslash_L)^i_j = \g^\m D^i_{\m j}L
\ee
and $ D^i_{\m j}$ given in (\ref{frstuff}). The relation (\ref{d2})
holds with field strength (see also \cite{Gaillard:1993es})
\bea \lab{fmn2}
\cf^i_{\m\n j} &=& R_{AB}{}^i{}_j D_\m z^A D_\n z^B - \mop Fa\!_{\m\n}\mop Xa\!^{i};_{j}
 +\half i B_{\m\n}\d^i_j\ ,
\non
R_{AB}{}^i{}_j &=& \O^i_{A j,B} + \O^i_{A l}\O^l_{B j} - \O^i_{B j,A} -\O^i_{B l}\O^l_{A j}\ .
\eea
It then follows from sections  6.4.1 and 6.4.2 of \cite{fuji} that the
 anomalous Jacobian $J(\a^aT^a)$ of the path integral measure is
\be \lab{jac}
\ln J(\mop \alpha a \mop Ta) = \frac{i}{32\pi^2}\int d^4x \, \mop \a a\!(x) \e^{\m\n\r\s}
\tr(\mop Ta \cf_{\m\n}\cf_{\r\s})\ .
\ee
The anomalous conservation law of the $U(n)$ current is
\bea \lab{janom}
\langle D_\n (\pb_i \mop Ta\!^{i}_j \g^\m L\psi^j)\rangle
~=~ \frac{i}{32\pi^2}  \e^{\m\n\r\s}
\tr(\mop Ta \cf_{\m\n}\cf_{\r\s})\ .
\eea
We now relate this result to the anomalous divergence in (\ref{con4}) by
expressing that current in the frame basis and using the
completeness relation of the matrices $T^a$, namely
\be
\mop Ta\!^{i}_j \mop Ta\!^{k}_l = \d^i_l\d^k_j\ ,
\ee
to write
\bea
\langle D_\n(\pb_{\b}\g^\n L\psi^\a) \rangle &=& e^j_\b e^\a_i \mop Ta\!^{i}_j
\langle D_\n(\pb \mop Ta \g^\n \psi)\rangle \non
&=&  \frac{i}{32\pi^2} e^j_\b e^\a_i \mop Ta\!^{i}_j \mop Ta\!^k_l
  \e^{\m\n\r\s}\cf^l_{\m\n m} \cf^m_{\r\s k}\non
&=&  \frac{i}{32\pi^2} \e^{\m\n\r\s}\cf^\a_{\m\n \g}\cf^\g_{\r\s \b}\ .
\eea
In the last line we converted the field strength to the coordinate basis
of $\ct$ in which
\be \lab{fs}
\cf^\a_{\m\n \b} = R_{\g\bd}{}^\a{}_\b
 (D_\m z^\g D_\n z^{\bd}-D_\n z^\g D_\m z^{\bd})- \mop Fa\!_{\m\n}\mop Xa\!^\a;_{\b}
+ \frac{i}{2} B_{\m\n}\d^\a_\b\ .
\ee
The inconsistency condition then reads
\be \lab{incon2}
0 = \langle D_\n \mop Ja\!^{\n} \rangle = -\frac{1}{32\pi^2}\Big[ G^{\b\bg} \mop Ya\!_{\a\bg} -
\frac{1}{2}{\rm Im}(\mop Fa) \d^\b_\a\Big] \e^{\m\n\r\s}\cf^\a_{\m\n
\g}\cf^\g_{\r\s \b}\ ,
\ee
in which we use the abbreviation
\beqn \lab{abbr}
\mop Ya\!_{\a\bb} = \frac{1}{2i} \Big(  G_{\g\bb} \mop Xa\!^{\g},_{\a}
- G_{\a\bg} \mop Xa\!^{\bg},_{\bb} \Big) \ .
\eeqn
When the field strength (\ref{fs}) is inserted in (\ref{incon2})
one finds a rather complex but correct expression for the anomaly.
Even when Im$(F^a)=0$ the anomaly contains new terms due to the
presence of $B_{\m\n}$ in (\ref{fs}).

%%%%%%%%%%%%%%%%%%%%%%%%%%%%%%%%%%%%%%%%%%%%%%%%%%%%%%%%%%%%%%%%%%%%%%%

\subsection{An example: $\mbb{CP}^1$}

It is useful to treat a specific example which has the features of
the general $\s$-model discussed in section  3.3, yet is simple
enough that one can obtain the consistency condition without the
full geometrical baggage. Therefore we outline briefly the case of
the target space $\mbb{CP}^1$. The isometry group is $SU(2)$ with
three holomorphic Killing vectors $X^{a\a}(z)$, $a=1,2,3$. The
K\"ahler potential, however, is not invariant under $SU(2)$. Thus
Im($F^a(z))\neq 0$, and K\"ahler anomalies play a role in the
consistency conditions. To analyze this model we need the K\"ahler
potential, metric, and connection
\beqn
K = \ln(1+z\bar z)\ , \quad
G_{z\bar z} = (1+z\bar z)^{-2}\ , \quad
\G^z_{zz} =-2\bar{z}(1+z\bar z)^{-1}\ ,
\eeqn
and the Killing vectors and D-terms
\beqn
&& \mop X1\!^z ~=~ -\frac{i}{2} (1-z^2) \ , \quad
\mop D1 ~=~ \frac12 \frac{z+\bar z}{1+z\bar z} \ , \non
&& \mop X2\!^z ~=~ \frac{1}{2} (1+z^2) \ , \quad ~~~
\mop D2 ~=~ -\frac{i}{2} \frac{z-\bar z}{1+z\bar z} \ , \non
&& \mop X3\!^z ~=~ -iz \ , \quad \quad \quad \quad
\mop D3 ~=~ -\frac{1}{2} \frac{1-z\bar z}{1+z\bar z} \ .
\eeqn
This leads to
\beqn
\mop F1 = \frac{i}{2} z\ , \quad
\mop F2 = \frac12 z \ , \quad
\mop F3 = -\frac{i}{2} \ .
\eeqn
Note that the K\"ahler potential is in fact invariant under the third
isometry $X^{3z}$, but still Im$(F^3)\neq 0$. Due to (\ref{dadj}) it
is fixed to the non-vanishing constant value above.

We consider the $\mbb{CP}^1$ $\s$-model, with a chiral fermion,
and with the full $SU(2)$ symmetry group gauged. However, to
simplify the equations, we set the gauge fields $A^1_\m$ and
$A^2_\m$ to zero\footnote{If we were to gauge only the $U(1)$
referring to $X^{3z}$ then the constant Im$(F^3)$ would be
arbitrary and interpreted as the  \FI coupling. In particular we
would have the freedom to set $F^3=0$, since the K\"ahler
potential is invariant.} and focus on the inconsistency related to
$X^{3z}$. Similar results would be obtained for the other two
isometries. We also leave out the gauginos. In this model the
Lagrangian (\ref{cfml}) becomes:
\bea \lab{cpl}
\cl &=& -\frac{1}{4} \mop F3\!_{\m\n}\mop F3\!^{\m\n} - \frac{1}{(1 +
z\bar{z})^2} \Big[(\pa_\m
  -i \mop A3\!_\m)\bar{z} (\pa^\m +i\mop A3\!^\m)z \non
&& \hspace{4cm}
- \pb \g^\m\Big( \pa_\m -\frac{2\bar{z}\pa_\m z}{1 + z\bar{z}} +
i\frac{1-z\bar{z}}{1 + z\bar{z}}\mop A3\!_\m +\half i B_\m\Big)
L\psi\Big]\ ,
\non
B_\m &=& \frac{{\rm Im}(\bar{z}\pa_\m z) - \half{(1-z\bar z)}
A^3_\m}{1+z\bar z}\ .
\eea
It is now straightforward to obtain by direct calculation, ignoring the
geometrical origin of the terms in (\ref{cpl}), the consistency condition
\beqn \lab{con7}
0 ~=~ - D_\mu \frac{\delta \cl}{\delta A_\mu^3}
   -iz \frac{\delta \cl}{\delta z} + i\bar z \frac{\delta \cl}{\delta
\bar z} &=&
\non
&& \hspace{-3cm}
- \frac{i}{(1+z\bar z)^2} D_\n (\bar\psi \gamma^\n L \psi)
- \frac{1}{4} \partial_\n\Big[ - \frac{i}{(1+z\bar z)^2} \bar\psi
\gamma^\n L \psi \Big] \ ,
\eeqn
where $D_\mu$ is the full $\sigma$-model covariant derivative. This
result can be compared with the general expression (\ref{incon2}).
The first term reproduces $Y_{\a\bb}^3 = - G_{\a\bb}$, the second
Im($F^3) = -{1}/{2}$. In this
simple model the two terms in the last line of (\ref{con7}) are
proportional, so
we get
\be \lab{con8}
0 = - \frac{3}{4} \pa_\n N^\n \ ,
\ee
as a special case of (\ref{incon2}), with the $U(1)$ axial current
\be \lab{ax3}
N^\n =-\frac{i}{(1+z\bar z)^2} \bar\psi \gamma^\n L \psi\ .
\ee
The axial anomaly is
\be \lab{cp1anom}
\langle \pa_\n N^\n \rangle =
\frac{1}{32\pi^2} \epsilon^{\m\n\r\s} \cf_{\m\n} \cf_{\r\s}\ ,
\ee
in which $\cf_{\m\n}$ is the field strength of the connection in
(\ref{cpl}), namely
\be \lab{fs3}
\cf_{\m\n} = \frac{2}{(1+z\bar{z})^2}(D_\m {z}D_\n \bar z -D_\n {z}D_\m \bar z)
+i \frac{1-z\bar{z}}{1+z\bar{z}} \mop F3\!_{\m\n} +\half i B_{\m\n}\ .
\ee
This agrees with (\ref{fs}) in the $\mbb{CP}^1$ model.

%%%%%%%%%%%%%%%%%%%%%%%%%%%%%%%%%%%%%%%%%%%%%%%%%%%%%%%%%%%%%%%%%%%%%%%

\section{The general gauged supergravity model}
\setcounter{equation}{0}
\lab{full}

The supersymmetric $\sigma$-model coupled to supergravity includes
the gravitino and various coupling terms in addition to the terms
studied in the previous sections. The action is
\be
\cs[e^i_\m,\mop Aa\!_\m,z^\a,z^{\bb},\psi^\a,\pb^{\bb},\mop \l a,\Psi_\m]=\int
d^4x \, {\rm det}(e^i_\m)\cl_{\rm SG}\ ,
\ee
with  Lagrangian density (in conventions similar to those of \cite{wb})
\beqn \lab{lsg}
\cl_{\rm SG} = \cl_b + \cl_f + \cl_{\rm int} +\, {\rm quartic\ terms}
\eeqn
with
\beqn
\cl_b &=& \frac12 R - \frac14 \mathop{F}^a{}_{\!\!\! \mu\nu}
\mathop{F}^a{}^{\!\mu\nu}
 - \frac12 \mop Da\!\! \mop Da
 - G_{\alpha\bar\beta} D_\mu z^\alpha D^\mu z^{\bar\beta} \ ,
\\
\cl_f &=& \frac12 \bar\Psi_\mu \gamma^{\mu\nu\rho} D_\nu \Psi_\rho
 + \frac12 \mop{\bar\lambda}a \gamma^\mu D_\mu \mop \lambda a
 + G_{\a\bb} \bar{\psi}^{\bb} \g^\m D_\m L \psi^\a \ ,
\non
\cl_{\rm int} &=&
\frac{1}{\sqrt 2} G_{\alpha\bar\beta} \big[ D_\mu z^{\bar\beta} \bar\Psi_\nu
 \gamma^\mu\gamma^\nu L \psi^\alpha + D_\mu z^{\alpha} \bar \psi^{\bar\beta}
 R \gamma^\nu\gamma^\mu \Psi_\nu \big] \non
&&
+\frac12 \mop Da \bar\Psi_\mu \gamma^\mu \gamma_5 \mop \lambda a
+\mop Fa _{\!\!\! \rho\sigma} \bar\Psi_\mu \gamma^{\rho\sigma}\gamma^\mu \mop \lambda a
+\sqrt 2 G_{\alpha\bar\beta} \big[ \mop{X}a ^{\!\! \bar\beta}
 \mop{\bar\lambda}a L \psi^\alpha + \mop {X}a ^{\!\! \alpha} \bar\psi^{\bar\beta} R \mop \lambda a \big]\ .
\nonumber
\eeqn
We omit the complicated set of four-fermion terms, see \cite{wb},
but our argument includes their effects, see also the footnotes
\ref{foot1} and \ref{foot2}.  We assume there is no
superpotential and minimal (i.e.\ field independent) gauge kinetic
functions to simplify the discussion. The gravitino covariant
derivative is defined as
\beqn \lab{dino}
D_\mu \Psi_\nu &=& \Big( \nabla_\mu + \frac12 iB_\mu \gamma_5
\Big) \Psi_\nu ~=~ \Big( \pa_\mu + \frac14 \omega_{\mu
ij}\gamma^{ij} + \frac12 iB_\mu \gamma_5 \Big) \Psi_\nu
\ ,
\eeqn
in which $\nabla_\mu$  includes the spin connection. Covariant
derivatives of the matter fields were given previously in
(\ref{dbos},~\ref{dlam},~\ref{dpsi2}). One must replace $\pa_\mu
\longrightarrow \nabla_\mu$ in (\ref{dpsi2}) and (\ref{dlam}).
Note that the composite K\"ahler connection (\ref{kcon}) couples
to all fermions.

The model has a global $U(1)_R$ axial symmetry with transformations
\be \lab{ax}
\d L\psi^\a = i\a L\psi^\a\ ,\quad
\d\mop \l a = i\a\g_5\mop \l a\ ,\quad
\d\Psi_\m = i\a\g_5\Psi_\m
\ee
and Noether current
\be \lab{noe3}
N^\m = - \frac{i}{2}\Big[ 2G_{\a\bb}\pb^{\bb}\g^\m L\psi^\a +\mop
{\bar{\l}}a\g^\m\g_5\mop \l a +
\bar{\Psi}_\r\g^{\r\m\n}\g_5\Psi_\n\Big]\ .
\ee
It is an $R$-symmetry since $z^\a$ is neutral while $L\psi^\a$,~
$L\l^a$, and $L\Psi_\m$ have charges $-1, +1,+1$, respectively.
The $U(1)_R$ symmetry is effectively gauged by $B_\m$. There is
also a gauge symmetry with parameters $\theta^a(x)$ with $\d
A^a_\m = D_\m \theta^a$. For the gauge variation of other fields
we use the notation $\d =\theta^a \d^a$. We then have
\bea \lab{gv}
\mop \d a z^\a &=& \mop Xa\!^\a\ ,\quad
\mop \d a z^{\bb} ~=~ \mop Xa\!^{\bb}\ , \non
\mop \d a L\psi^\a &=& \mop Xa\!^{\a},_{\b}L\psi^\b- \frac{i}{2}{\rm Im}(\mop Fa) L\psi^\a\ ,\quad
\mop \d a \pb^{\bb}R ~=~ \mop Xa\!^{\bb},_{\bg}\pb^{\bg}R +\frac{i}{2}{\rm Im}(\mop Fa) \pb^{\bb}R\ , \non
\mop \d b \mop \l a &=& - f^{abc}\l^c -\frac{i}{2} {\rm Im}(\mop Fb) \g_5 \l^a\ ,\quad
\mop \d a \Psi_\m ~=~  -\frac{i}{2} {\rm Im}(\mop Fa) \g_5\Psi_\m\ .
\eea
Holomorphic Killing vectors $X^{a\a}(z),~X^{a\bb}(\bar{z})$ and
the holomorphic function $F^a(z)$ induced by a gauge
transformation of the K\"ahler potential were discussed in section
2. The gauge invariance of the theory is expressed by the identity
\bea \lab{gi}
\d \cl_{\rm SG} ~=~ 0
&=& \mop \theta a(x)\Big[
-D_\n \frac{\d \cl_{\rm SG}}{\d A^a_\n} +\mop Xa\!^\a \frac{\d \cl_{\rm SG}}{\d
z^\a}
+\mop Xa\!^{\bb}\frac{\d \cl_{\rm SG}}{\d z^{\bb}}\\
&& \hspace{1cm}
+ \mop \d a \pb^{\bb}R\frac{\d \cl_{\rm SG}}{\d \pb^{\bb}}+
\frac{\d \cl_{\rm SG}}{\d \psi^\a} \mop \d a L\psi^\a + \mop \d a \mop {\bar{\l}}b
\frac{\d \cl_{\rm SG}}{\d \bar{\l}^b} +
\mop \d a \bar{\Psi}_\r \frac{\d \cl_{\rm SG}}{\d \bar{\Psi}_\r}\Big]
\nonumber
\eea
which is the same as (\ref{gi1}) applied to the general supergravity
Lagrangian. The gauge field equation of the model reads
\bea \lab{gfe}
D_\m \mop Fa\!^{\m\n} &=& \mop Ja\!^{\n} ~\equiv~ -\frac{\d
(\cl+\frac14 F^b_{\mu\nu} F^{b\mu\nu})}{\d A^a_\n}
\non
&=& \mop Ja\!^{\n}_b + \mop Ja\!^{\n}_f + \mop ja\!^{\n} + \half \mop Da N^\n + \mop Ja\!^{\n}_{\rm int}\ .
\eea
with $J^a_b$ and $J^a_f$ defined
in (\ref{aeq2}), $j^{a\n} = \frac12 f^{abc}\bar{\l}^b \g^\n \l^c$ and
\bea \lab{jint}
\mop Ja\!^{\n}_{\rm int} &=& -\frac{\d \cl_{\rm int}}{\d A^a_\n}\non
&=& \frac{1}{\sqrt{2}} G_{\a\bb} \Big[ \mop Xa\!^{\bb} \bar{\Psi}_\r\g^\n\g^\r L\psi^\a
 + \mop Xa\!^{\a} \pb^{\bb}R\g^\r\g^\n\Psi_\r\Big] +
 2 D_\m(\bar{\Psi}_\r \g^{\m\n}\g^\r \mop \l a)\ .
\eea
To derive the consistency condition we now follow the same
strategy as above. Assuming that the gauge variation of the action
from varying bosons vanishes by the scalar equations of motion,
the consistency condition arises from the fermion variations. The
supergravity generalization of the expressions obtained earlier
for only gauginos in
(\ref{anom1}) and for only chiral fermions in (\ref{con4}) turns
out to be
\be \lab{con3}
0= \langle \nabla_\n \mop Ja^{\n} \rangle =  i \mop Ya\!_{\a\bb} \langle \nabla_\n(\pb^{\bb}\g^\n
L\psi^\a)\rangle  + \< \nabla_\m \mop ja^\m \>
+ \half {\rm Im}(\mop Fa) \langle \nabla_\n
N^\n\rangle
\ ,
\ee
in which the $\nabla_\n$ derivative carries appropriate
space-time, target space and gauge connections, and $\langle ...
\rangle$ again indicates just the anomalous divergences of the
currents. Comparing with (\ref{gfe}), we have dropped the
divergence of $J^{a\n}_{\rm int}$. As we argue in the appendix,
this does not affect the anomaly. The condition (\ref{con3}) is
the central result of our analysis.

The proper K\"ahler anomaly, proportional to Im$(F^a)$, is the
third term of (\ref{con3}). The anomaly has contributions from
gauginos as in (\ref{axanom}), from chiral fermions, as can be
inferred from (\ref{incon2}), and from the gravitino. The
gravitino gauge anomaly is 3 times that of a gaugino, but coupled
only through the K\"ahler connection in (\ref{dino}). We obtain
\be \lab{nanom}
\langle \nabla_\n N^\n\rangle_{\rm gauge} = \frac{1}{32\pi^2}\e^{\m\n\r\s}
\Big[ C_2(G)\mop Fa\!_{\m\n}\mop Fa\!_{\r\s} +\frac{n_\l + 3}{4}B_{\m\n}B_{\m\n} +
C_{\m\n\r\s}\Big] \ .
\ee
To write the contribution of the chiral fermions we first define
\beqn
\Sigma_{\m\n\b}^\a &=& R_{\g\bd}{\,\!}^\a{\,\!}_\b
 (D_\m z^\g D_\n z^{\bd}-D_\n z^\g D_\m z^{\bd})\ ,
\eeqn
which is essentially the target space curvature tensor pulled back to
spacetime. Using this we form 
\beqn \lab{mess}
C_{\m\n\r\s} &=& \Sigma_{\m\n\b}^\a \Sigma_{\r\s\a}^\b
 + \mop Fa\!_{\m\n} \mop Fb\!_{\r\s} \mop Xa\!^\a;_\b \mop Xb\!^\b;_\a
 - \frac14 n_\psi B_{\m\n} B_{\r\s}  \non
&&
 - i \mop Fa\!_{\m\n} B_{\r\s} \mop Xa\!^\a;_\a
 - 2 \mop Fa\!_{\m\n} \Sigma_{\r\s\a}^\b \mop Xa\!^\a;_\b +i \Sigma_{\m\n\a}^\a B_{\r\s}\ .
\eeqn

The anomaly of the gaugino current $j^{a\m}$ in the second term of
(\ref{con3}) is identical to the truncated model of section \ref{secgaug} given
in (\ref{gauganom}). The contribution of the chiral fermions to
the first term in (\ref{con3}) is
\beqn \lab{messier}
\hspace{-.5cm}
\mop Ya\!_{\a\bb} \langle\nabla_\n(\pb^{\bb}\g^\n L\psi^\a)\rangle_{\rm gauge} &=&
\frac{i}{32\pi^2}\e^{\m\n\r\s}
G^{\b\bar\delta} \mop Ya\!_{\a\bar\delta}
\Big[ \Sigma_{\m\n\g}^\a \Sigma_{\r\s\b}^\g 
 + \mop Fb\!_{\m\n} \mop Fc\!_{\r\s} \mop Xb\!^\a;_\g \mop Xc\!^\g;_\b
\\ 
&&
\hspace{-4.0cm}
 - \frac14 B_{\m\n} B_{\r\s} \d_\b^\a
 - i \mop Fb\!_{\m\n} B_{\r\s} \mop Xb\!^\a;_\b
 + i \Sigma_{\m\n\b}^\a B_{\r\s} 
%\non
%&&
%\hspace{-1cm}
 - \mop Fb\!_{\m\n} \Big( \Sigma_{\r\s\g}^\a \mop Xb\!^\g;_\b + \Sigma_{\r\s\b}^\g \mop Xb\!^\a;_\g \Big)
   \Big] \ .
\nonumber 
\eeqn
In the case of a flat target space $\ct = \mathbb C^{n_\psi}$ and
a linear realization of gauge symmetry, the Killing vector
derivative reduces to constants, $X^{a\a};_\b = X^{a\a},_\b \to
T^{a\a}{}_\b$, a matrix generator of the gauge group $G$. In this
case the second term of (\ref{messier}) reduces to the
conventional cubic gauge anomaly of the chiral fermions.

The gravitational anomaly is more conventional. See Chapter 10 of
\cite{fuji}, for example, for spin $\frac12$ fields.
The contribution of the gravitino to the anomaly of the
Noether current is $-21$ times that of a gaugino. The gaugino current
$j^{a\m}$ itself has no gravitational anomaly. The complete result is given by
\bea \lab{gravanom}
\langle \nabla_\n N^\n\rangle_{\rm grav} &=& -\frac{1}{768\pi^2}(n_\l -21 - n_\psi)
\, \epsilon^{\m\n\r\s} R_{\m\n\xi\t}R_{\r\s}{}^{\xi\t}\ ,
\non
\mop Ya\!_{\a\bb} \langle\nabla_\n(\pb^{\bb}\g^\n L\psi^\a)\rangle_{\rm grav}&=& - \frac{i}{768\pi^2}
\mop Ya\!_{\a\bb} G^{\a\bb}
\, \epsilon^{\m\n\r\s} R_{\m\n\xi\t}R_{\r\s}{}^{\xi\t}\ .
\eea
One can see that the gauge anomaly is very complicated. 
As a general observation, it is not possible
to cancel the coefficient $n_\l +3 -n_\psi$ of $B_{\m\n}B_{\r\s}$
in (\ref{nanom}) and (\ref{mess}) for the gauge anomaly and the
gravitational anomaly in (\ref{gravanom}) at the same time by
adjusting $n_\l$ and $n_\psi$.

If ${\rm Im}(F^a)=0$, then the
$\pa_\n N^\n$ anomaly is absent. However, there are still several
new terms involving $B_{\m\n}$ which can affect the consistency of the
model. Anomaly cancellation will be studied in \cite{EFK} with emphasis
on the case of a flat target space. 

%%%%%%%%%%%%%%%%%%%%%%%%%%%%%%%%%%%%%%%%%%%%%%%%%%%%%%%%%%%%%%%%%%%%%%%%

\section{Flux compactifications: gauged shift symmetries}
\setcounter{equation}{0}
\lab{GL}

In this section we apply the general formalism developed earlier
to a supergravity model with a gauged shift symmetry. This form of
gauge symmetry arises naturally in compactifications of
ten-dimensional supergravity or string theory with background
fluxes for the $p$-form field strengths along the internal
directions \cite{Grana:2005jc}. Specifically, we will use a
truncation of the $\cn=1$ flux vacuum model found in
\cite{Grimm:2004uq}. Similar structures also occur in models with
gauged $\cn=4$ supersymmetry \cite{D'Auria:2002tc}. In the latter
case one deals with a toroidal flux compactification. The gauging
of shift symmetries is evident from the dimensional reduction of
the type IIB 5-form, schematically\footnote{The indices are $M,\,
N,...$ for ten dimensions, $m,\, n,...$ for internal six
dimensions and $\m,\, \n,...$ for four dimensions. The $C^{(p)}$
are RR $p$-forms with field strengths $F^{(p+1)}$, $B^{(2)}$ the
NSNS 2-form with field strength $H^{(3)}$.}
\beqn
\frac15 F^{(5)}_{MNOPR} = \pa_{\m} C^{(4)}_{nopr} +
2 C^{(2)}_{\m[n}H^{(3)}_{opr]} - 2 B^{(2)}_{\m[n}F^{(3)}_{opr]} \
.
\eeqn
The kinetic term for the scalars $C^{(4)}_{nopr}$ in the
four-dimensional Lagrangian  then contains a coupling to the
vector bosons $C^{(2)}_{\m n}$ and $B^{(2)}_{\m n}$, and the
coupling constants are given by the values of the 3-form fluxes
$F^{(3)}_{opr}$ and $H^{(3)}_{opr}$. The vectors thus gauge the
shift symmetries of the scalars whenever fluxes are present.

In $\cn=1$ supersymmetric Calabi-Yau flux compactifications a
similar gauging arises. The truncated model we will discuss 
includes an abelian vector multiplet with fields $(A_\mu,\lambda)$
and one chiral multiplet with $(L\psi, S=e^\phi
+ ih)$. In
\cite{Grimm:2004uq} there are several abelian gauge fields
$A_\m^i$ which arise from the reduction of the RR 4-form along
3-cycles of a Calabi-Yau manifold. They couple to scalars of
chiral multiplets with charges $e_i$ determined by flux quantum
numbers. The scalar $S$ which we retain in our truncation is the
universal scalar of the string compactification. It involves the
string coupling and the RR axion $h$. It parameterizes the well
known non-compact $SU(1,1)/U(1)$ manifold. Scalars and vectors
couple through the covariant derivative $D_\mu
S =
\pa_\mu e^\phi
+ i (\pa_\mu h- e A_\mu)$. The gauge symmetry is thus the shift symmetry
\beqn
\delta h = \theta(x) \ , \quad \d \phi =0\ , \quad
\delta A_\mu = \frac{1}{e} \partial_\mu \theta(x)\ .
\eeqn
The corresponding Killing vectors are imaginary constants and read
\beqn
X^S = - X^{\bar S} = ie\ .
\eeqn
The K\"ahler potential, which is gauge invariant in this model, is
\beqn\label{kaehler}
K = -\ln(S+\bar S)\ .
\eeqn
The abelian gaugino $\l$ is always gauge invariant, and  the chiral
fermion $\psi$
is gauge invariant in this model since $X^S,_S=0$.
The $D$ field is  defined by the differential equation
\beqn
G_{S\bar S} X^{\bar S} = i D,_S\ , \quad
G_{S\bar S} X^{S} = -i D,_{\bar S}\ ,
\eeqn
with general solution
\beqn\lab{dfield}
D = \frac{e}{S+\bar S} + \xi \ .
\eeqn
The explicit dimensional reduction of \cite{Grimm:2004uq} leads to
a potential energy from D-terms given by
\beqn\label{dterm}
V = \frac18 e^2 e^{-2\phi} = \frac12 D^2 \ ,
\eeqn
which implies that $\xi=0$, even though a non-vanishing value would
have been compatible with $D=4$ supergravity.

The full Lagrangian of the model is a special case of (\ref{lsg}) and
rather simple. It reads
\beqn
\cl &=& \cl_b + \cl_f + \cl_{\rm int} \ , \\
\cl_b &=& \frac12 R - \frac14 F_{\mu\nu}F^{\mu\nu}
- G_{S\bar S} D_\mu S D^\mu \bar S - \frac12 D^2  \ ,
\non
\cl_f &=& \frac12 \epsilon^{\mu\nu\rho\sigma} \bar\Psi_\mu \gamma_\nu
D_\rho \Psi_\sigma
  + \frac12 \bar\lambda \gamma^\mu D_\mu \lambda
  + G_{S\bar S} \bar{\psi}\g^\m D_\m L \psi \ ,
\non
\cl_{\rm int} &=&
\frac{1}{\sqrt 2} G_{S\bar S} \big[ D_\mu \bar S
\bar\Psi_\nu\gamma^\mu\gamma^\nu L \psi
  + D_\mu S \bar \psi R \gamma^\nu\gamma^\mu \Psi_\nu \big] \non
&&
+\frac12 D \bar\Psi_\mu \gamma^\mu \gamma_5 \lambda
+ F_{\rho\sigma} \bar\Psi_\mu \gamma^{\rho\sigma}\gamma^\mu \lambda
+\sqrt 2 G_{S\bar S} \big[ X^{\bar S} {\bar\lambda} L \psi +
   X^{S} \bar\psi R \lambda \big] \ ,
\nonumber
\eeqn
where
\beqn
D_\mu \Psi_\nu &=& \Big( \nabla_\mu + \frac12 i B_\mu \gamma_5 \Big)
\Psi_\nu \ , \non
D_\mu  \lambda &=& \Big( \nabla_\mu + \frac12 i B_\mu \gamma_5
\Big)\lambda \ , \non
D_\mu L \psi &=& \Big( \nabla_\mu + \Gamma^S_{SS} D_\mu S + \frac12 i
B_\mu \Big)
L \psi \ .
\eeqn
The composite K\"ahler connection is 
\beqn
B_\mu = \frac{1}{2i} \Big( K,_S D_\mu S - K,_{\bar S} D_\mu \bar S
\Big) = - \frac{1}{S+\bar S} (\pa_\mu h - e A_\mu ) \ . 
\eeqn
It is gauge invariant in this model because the K\"ahler potential
is invariant and Im$(F)=\xi=0$. One can now directly obtain the equations of
motions for $A_\mu$ and $h$ (without going through those of $S$
and $\bar S$ first), which are
\beqn
\nabla_\mu F^{\mu\nu} + 2 e G_{S\bar S} ( \partial^\nu h - e A^\nu)
&=& e J^\nu \ , \non
2 \nabla_\mu \big( G_{S\bar S} ( \partial^\mu h - e A^\mu)\big)
&=& \nabla_\mu J^\mu \ ,
\eeqn
where
\beqn
J^\mu = - \frac{\delta(\cl_f+\cl_{\rm int})}{\delta A_\mu}\ .
\eeqn
Applying $\frac{1}{e}\nabla_\nu$ one finds an expression which
vanishes completely when the scalar equation of motion is used.
Thus there is no inconsistency in this model. This agrees with the
general consistency condition (\ref{gi}) because we have a gauged shift
symmetry with constant Killing vectors and $B_\m$ is gauge invariant.
Therefore the fermions are invariant under the gauge transformation
and all terms in (\ref{gi}) cancel when the scalar equations of motion
are used. Hence no inconsistency can arise.

This favorable result depends on the particular choice of K\"ahler
potential (\ref{kaehler}) which is gauge invariant under the shift symmetry.
A general K\"ahler transformation would lead to a non-gauge invariant
potential and thus a non-vanishing Im$(F)$. By (\ref{con3}) this would signal that the
theory is inconsistent. The resolution of this 
apparent problem is that one need not require K\"ahler invariance
in this model because the target space is topologically trivial
and the K\"ahler potential can be chosen to be gauge invariant.
There is no need to consider K\"ahler transformations which change
the preferred form (\ref{kaehler}).

In other situations the absence of invariance under K\"ahler
transformations can lead to severe problems. A class of examples are
toroidal orbifold compactifications where modular $SL(2,\mbb Z)$ transformations of
the background tori act on the moduli scalars as perturbatively exact global
symmetries. They leave the K\"ahler potential only invariant
up to K\"ahler transformations. The cancellation of anomalies restricts
the charged matter spectrum of these models, as was for example
studied in \cite{Ibanez:1992hc}. No such problem actually arises in
the present case. The symmetry $S\mapsto S'=-1/S$ of the target space 
$SU(1,1)/U(1)$ looks potentially dangerous since it 
leads to a K\"ahler potential not invariant under a shift of $S'$. 
However, the inversion is not a symmetry of the full model
because it is broken by the D-term potential (\ref{dterm}) induced by the
fluxes. Thus, the choice of flux breaks the global symmetry and no
inconsistency arises.\footnote{We would like to thank Jan Louis
for this key observation.}

The problem of K\"ahler anomalies can reemerge if one tries to
``integrate out'' the scalar $S$ replacing it with a constant
background value. Then the D-term of (\ref{dfield}) takes the role
of a constant \FI coupling, since $S$ is no longer dynamical. This
is the philosophy often adopted in the case of D-terms generated
in the context of the Green-Schwarz mechanism. The classic example
is the \FI coupling of Dine, Seiberg, and Witten
\cite{Dine:1987xk}. After fixing $S$ the issue of the consistency
may need to be readdressed.\footnote{The question if $S$ can
be integrated out without breaking supersymmetry was discussed in \cite{bdkvp}.}

It is curious that if the gravitational degrees of freedom in the model
of this section are dropped and the field Re$(S) = e^\phi$ is frozen at a constant value,
the model is essentially the same as that considered in section 2 of
\cite{gj}. Gross and Jackiw found that the model has an axial anomaly
but is consistent. It is a model of a massive vector boson
in which the anomaly induces a non-renormalizable term
$h\, \epsilon^{\m\n\r\s} F^a_{\m\n} F^a_{\r\s}$.
This shows that there are models with triangle anomalies which
are nevertheless consistent.

%%%%%%%%%%%%%%%%%%%%%%%%%%%%%%%%%%%%%%%%%%%%%%%%%%%%%%%%%%%%%%%%%%%%%%%%%%%%%%%%%%%%%

\section{Discussion and Conclusions}
\setcounter{equation}{0}
\lab{discussion}

To summarize, we have shown that the structure of anomalies in
gauged non-linear $\s$-models coupled to supergravity is richer
and more intricate than often assumed in the literature. This is
mainly due to the fact that all fermions couple to the composite
K\"ahler gauge connection $B_\mu$. This $B_\m$ and its field
strength $B_{\m\n}$ depend on the scalar fields. In addition to
the usual gauge, gravitational, and $\s$-model anomalies,
$B_{\m\n}$ appears in the violation of gauge current conservation
laws via one-loop triangle diagrams which threatens to spoil the
consistency of the theory.

The K\"ahler transformation $K(z,\bar z) \to K(z,\bar z) + f(z) + \bar{f}(\bar z)$
is at the root of the anomaly and consistency issue. However,
invariance under K\"ahler transformations is not necessarily required in a field theory model.
We now clarify the conditions under which a significant consistency
problem occurs.
\begin{enumerate}
\item Suppose that the $\s$-model target space $\ct$ is topologically trivial,
and there is a K\"ahler potential which is invariant under the
global and gauge symmetries of the theory. Then there is a priori
no need to go beyond this ``preferred" $K(z,\bar z)$, and the
theory is consistent. This was the case in the flux model of
section \ref{GL} where only shift symmetries were gauged.
\item If $\ct$ is topologically non-trivial, then several coordinate
charts ${\cal O}_A$ are required to cover $\ct$. The K\"ahler
potential need not be a globally defined 
scalar on $\ct$; rather there is a $K_A$ on each chart such that in overlap regions
$\co_A \cap\co_B$, the potentials are related by $K_A - K_B = F_{AB}$, where $F_{AB}$
is holomorphic. The spaces $\mbb{CP}^n$ are examples. In this case invariance under
K\"ahler transformation does impose consistency conditions
on the supergravity model. Witten and Bagger \cite{bagwit} 
discussed important constraints even in the absence of gauging
of the isometries. In addition the K\"ahler anomalies associated
with the gauging which we have emphasized are also significant. 
\item The two main situations analyzed in our paper are when  $K(z,\bar{z})$ is not
gauge invariant, but changes by a K\"ahler transformation as in
(\ref{ktrf2}), and the case of a \FI coupling. In both cases the
consistency condition (\ref{con3}) of the theory contains the extra term
Im$(F^a)\nabla_\n N^\n$ from (\ref{nanom}) where $N^\n$ is the Noether current of the
global axial symmetry. This is a challenge to the consistency of
all gauged supergravity theories where no gauge invariant K\"ahler potential
exists, such as $\mbb{CP}^n$, and to many phenomenological models
that make use of \FI couplings. Even 
when Im$(F^a) =0$ the consistency condition contains several new terms
due to $B_{\m\n}$ in (\ref{messier}) which must eventually be
canceled.   
\end{enumerate}
Conventional anomalies involve $\e^{\m\n\r\s}F_{\m\n}F_{\r\s}$ and
$\e^{\m\n\r\s} R_{\m\n\l\t}R_{\r\s}{}^{\l\t}$. 
These structures appear in gauge current anomalies and lead
to inconsistency unless cancelled. It is common practice to
attempt to cancel them by adding new fermions to the model. We
have found new anomaly structures which involve scalar fields, and
these can require new independent cancellation conditions.

It is well known that to cancel one-loop anomalies it is possible
to incorporate additional terms into a supergravity Lagrangian.
This has been widely investigated in string compactifications
where anomaly cancellation in the effective four-dimensional
theory is very important. See e.g.\
\cite{Moore:1984dc,LopesCardoso:1991zt,Derendinger:1991kr,Gaillard:1993es,Kaplunovsky:1994fg,ibanez,serone}.
Let us briefly recall some aspects of the known mechanisms which
have been studied. A supergravity theory may contain non-minimal
 field-dependent gauge kinetic functions $f_{ab}(z)$, which lead to the
term
\beqn
{\rm Im}(f_{ab}(z)) \epsilon^{\m\n\r\s} \mop Fa\!_{\m\n}\mop Fb\!_{\r\s}
\eeqn
in the action. If $f_{ab}(z)$ is not invariant under gauge transformations,
its variation contributes to the current
conservation in the same way as an anomalous triangle diagram. It 
can cancel terms in the gauge current anomalies in
(\ref{nanom}) and (\ref{messier}) which are
quadratic in $F^a_{\m\n}$. This is essentially a realization of
the four-dimensional Green-Schwarz mechanism. 

In \cite{EFK} the anomaly cancellation conditions of supergravity models with
gauge and K\"ahler anomalies, and with Green-Schwarz mechanism will be analyzed
in the limit of a flat $\s$-model target space. Essential steps to obtain
the physically relevant anomalies involve conversion of covariant
anomalies into consistent anomalies and including all finite local counter
terms in the Lagrangian. The outcome shows the necessity of a
Green-Schwarz mechanism whenever the K\"ahler potential is not gauge
invariant or \FI couplings are present.

Other possible counter terms based on superspace integrals were
proposed in
\cite{LopesCardoso:1991zt,Derendinger:1991kr} to cancel anomalies. These
terms are
non-local and of the form
\beqn
\epsilon^{\m\n\r\s} \mop Fa\!_{\m\n}\mop Fa\!_{\r\s} \frac{1}{\Box}
\nabla^\rho B_\rho\ , \quad
\epsilon^{\m\n\r\s} B_{\m\n}B_{\r\s} \frac{1}{\Box} \nabla^\rho B_\rho\ , ~~ ...
\eeqn
Their gauge variation is a local expression again of the same form as
induced by triangle anomalies with the respective gauge fields at the
vertices. Although these terms have been studied in several string
compactifications, we are not aware of any example where all anomaly structures
found in our work were demonstrated to cancel.

%%%%%%%%%%%%%%%%%%%%%%%%%%%%%%%%%%%%%%%%%%%%%%%%%%%%%%%%%%%%%%%%%%%%%%%%

%\clearpage
\begin{center}
{\bf Acknowledgements}
\end{center}
\vspace{-.3cm}

We would like to thank in particular Henriette Elvang for cooperation on 
various improvements of an earlier draft of this paper. 
Further we acknowledge helpful discussions and correspondence 
with Luis \'Alvarez-Gaum\'e, Jean-Pierre Derendinger, Sergio Ferrara,
Mary K.~Gaillard, Michael Haack, Matt Headrick, Arthur Hebecker,
Satoshi Iso, Roman Jackiw, Bj\"orn Lange, Jan Louis, Dieter
L\"ust, Michael Schulz, Angel Uranga, Giovanni Villadoro, and Fabio Zwirner. 
The research of D.~Z.~F.\
is supported by the NSF grant PHY-00-96515, and B.~K.\ enjoys the
support of the DFG, the DAAD, and the European RTN Program
MRTN-CT-2004-503369. Further support comes from funds provided by
the U.S. Department of Energy
(D.O.E.) under cooperative research agreement
$\#$DF-FC02-94ER40818.

\vspace{0cm}

%%%%%%%%%%%%%%%%%%%%%%%%%%%%%%%%%%%%%%%%%%%%%%%%%%%%%%%%%%%%%%%%%%%%%%%%%%%%%%%%%%%%%

\begin{appendix}

\section{Some arguments on one-loop anomalies}
\setcounter{equation}{0}

In the complete supergravity Lagrangian certain interaction terms
were neglected in the derivation of anomalies. Here we show that
there are no anomalous effect resulting from the current
$J^{a\n}_{\rm int}$ of
(\ref{jint}).

We begin by discussing the gauge-fixing of the gravitino action assuming a
flat space-time background for simplicity.
Let's add a gauge-fixing term to the gravitino
kinetic Lagrangian which includes the $B_\m$ connection, obtaining
\be \lab{lgf}
\cl = \frac12 (\bar\Psi_\mu \gamma^{\mu\nu\rho} D_\nu \Psi_\rho +\zeta
\bar\Psi_\mu \g^\m
\g^\n D_\n\g^\r\Psi_\rho)\ .
\ee
In the vNV gauge \cite{vnv}, in which $\zeta = -\frac12$, this
becomes
\be \lab{lgf2}
\cl = -\frac14 \bar\Psi_\mu \g^\r\g^\n D_\n\g^\m\Psi_\rho\ .
\ee
The linear field redefinition \cite{endo}
\be \lab{redef}
\Psi_\m \mapsto U_\m^\n \Psi_\n\ , \quad
U_\m^\n =\d_\m^\n -\frac12 \g_\m
\g^\n\ , \quad
U_\m^\r U_\r^\n =\d_\m^\n
\ee
takes us to the ultra-simple AGW gauge \cite{agw} Lagrangian
\be \lab{agwl}
\cl = \frac12 \bar\Psi^\mu \g^\n D_\n \Psi_\m \ .
\ee
Using standard $\g$-matrix identities the current can be written in terms
of the new variable $\Psi_\m$ as
\be \lab{jint2}
\mop Ja\!^\n_{\rm int} = \sqrt{2}G_{\a\bb}\Big( \mop Xa\!^{\bb}\bar{\Psi}^\n L\psi^\a
+\mop Xa\!^{\a}\pb^{\bb}R\Psi^\n\Big) + 2D_\m\Big(\bar\Psi_\r \g^{\m\n}\g^\r
\mop\l a \Big)\ .
\ee
Things simplify and there is little loss of generality if we study the
situation in the truncated Grimm-Louis model of section \ref{GL} in which $L\psi^\a$ can be
replaced by the single Majorana spinor $\psi$ and the total fermion current of the
model is
\bea \lab{glcur}
J^\n &=& - \frac i4 \frac{e}{S+\bar S} \Big(\bar{\l} \g^\n\g_5 \l
 + \bar\Psi^\m \g^\n\g_5 \Psi_\m
 - \frac{5}{(S+\bar S)^2} \bar\psi \g^\n \g_5 \psi \Big)
 + J^\n_{\rm int}\ ,
\non
J^\n_{\rm int} &=& \sqrt{2}\frac{ie}{(S+\bar{S})^2}(\bar\Psi^\n \g_5 \psi)
+2 \pa_\m(\bar\Psi_\r \g^{\m\n}\g^\r \l)\ .
\eea
We look for possibly anomalous triangle diagrams for the 3-point function
of currents $\< J^\m(z) J^\n(y) J^\r(y)\>$ in which the current $J^{a\m}_{\rm int}$ appears
in at least one position. Among diagrams with an internal $\psi$ line (and no
gauginos), one can rapidly see that the either Wick contractions vanish
or the diagrams have vanishing $\g$-matrix trace.

There are two non-vanishing one-loop diagrams involving two
insertions of the gaugino part of $J^{a\m}_{\rm int}$ and an insertion of
the gaugino and (gauge-fixed) gravitino axial currents. These
diagrams are complicated and a regulated calculation appears to be
difficult. Therefore we adopted another strategy, in which we
generalize the Fujikawa analysis to include the mixing of a
gravitino and (abelian) gaugino in the Lagrangian
\bea \lab{lmix}
\cl &=& \half \bar\Psi^\m \g^\n D_\n \Psi_\m + \half \bar{\l} \g^\n
D_\n\l  + \bar\Psi_\m \g^{\r\s}F_{\r\s} \g^\m \l\ ,
\eea
with $D_\n = \pa_\n + \half i B_\n \g_5$.
We then defined the operators $\cd_L= \cd L$ and $\cd_R = \cd R$ in
which $\cd$ is the matrix operator
\be  \lab{matop}
\cd = \left(
\begin{array}{cc}
\Dslash \delta_\mu^\nu & F_{\a\b}\g^{\a\b} \g_\mu \\
\g^\nu F_{\a\b}\g^{\a\b} & \Dslash
\end{array}
\right)\ .
\ee
It acts on $(\Psi_\nu, \lambda)^T$ to the right and on $(\bar\Psi^\mu, \bar\lambda)$
to the left. Mode expansions lead to a Jacobian in the Fujikawa method which is a
generalization of that of sections 6.4 and 6.4.1 of \cite{fuji} and reads
\be \lab{fujfac}
\ln(J) = -2i \int d^4k\, \a(x) e^{-ik\cdot x}
 {\rm tr}\Big[ f\Big(\frac{\cd\cd}{M^2}\Big) \g_5 \Big] e^{ik\cdot x}\ .
\ee
The function $f(\cd\cd/M^2)$ is a smoothly decreasing function of the
mode eigenvalues with cutoff scale $M^2$. After shifting $\Dslash$ by the
plane wave momentum $k_\m$ and expanding in powers of $1/M$ we
find that potentially non-vanishing terms of order up to
$1/M^4$ exist. In the standard calculation with $\Dslash$
instead of $\cd$ one only has to evaluate the product $\Dslash\Dslash$ and extract
the term $(\gamma^{\a\b}F_{\a\b})^2$ that has a non-vanishing trace with $\g_5$. Here it becomes
necessary to compute $\cd^4$ and consider many independent contributions.
The calculation is too tedious to report in detail.
Suffice it to say that, by careful evaluation of traces,
we were able to show that all effects of the $\bar\Psi_\m
\g^{\r\s}F_{\r\s} \g^\m \l$ mixing term vanish, and the only contribution to the
trace comes from the conventional terms $(\Dslash\Dslash)^2$. The gauge anomaly
reduces (after inclusion of gravitino ghosts) to the well known
anomaly of $\cn =1$ supergravity coupled to one gauge
multiplet.

\end{appendix}

%%%%%%%%%%%%%%%%%%%%%%%%%%%%%%%%%%%%%%%%%%%%%%%%%%%%%%%%%%%%%%%%%%%%%%%%

\end{document}